\documentclass[twocolumn]{article}
\usepackage[T1]{fontenc}
\usepackage{lmodern}
\usepackage[left=2.cm, right=2.cm, top=2.cm, bottom=2.cm]{geometry}
\usepackage{graphicx}
\usepackage{xcolor}
\usepackage{caption}
\usepackage[english]{babel}
\usepackage[super,comma,sort&compress]{natbib}
\usepackage[hidelinks]{hyperref}
\usepackage{amsmath}
\usepackage{setspace}
\usepackage{gensymb}
\usepackage{soul}
\usepackage[normalem]{ulem}
\usepackage{hyphenat}
\usepackage{siunitx}
\usepackage{times}
\usepackage{url}

\providecommand{\keywords}[1]
{
  \small	
  \textbf{\textit{Keywords: }} #1
}

\begin{document}
\renewcommand{\abstractname}{\vspace{-\baselineskip}}

\title{Unravelling the Band Structure and Orbital Character of a $\pi$-Conjugated \\ 2D Graphdiyne-Based Organometallic Network}
\author{Paolo D'Agosta$^{a}$, Simona Achilli$^{b,*}$, Francesco Tumino$^{a,c}$, Alessio Orbelli Biroli$^{d}$, Giovanni Di Santo$^{e}$, \\ Luca Petaccia$^{e}$, Giovanni Onida$^{b}$, Andrea Li Bassi$^{a}$, Jorge Lobo-Checa$^{f,g,*,\dag}$, and Carlo S. Casari$^{a,*,\dag}$}
\date{\begin{flushleft} \small $^{a}$Department of Energy, Politecnico di Milano, via G. Ponzio 34/3, I-20133 Milano, Italy \linebreak
$^{b}$Department of Physics ``Aldo Pontremoli’', Universit\`a degli Studi di Milano, Via G. Celoria 16, I-20133 Milano, Italy \linebreak
$^{c}$Department of Chemistry, Queen’s University, 90 Bader Lane, K7L3N6 Kingston, ON, Canada \linebreak
$^{d}$Department of Chemistry, Universit\`a di Pavia, Via Taramelli 12, I-27100 Pavia, Italy
\linebreak
$^{e}$Elettra Sincrotrone Trieste, Strada Statale 14 km 163.5, I-34149 Trieste, Italy \linebreak
$^{f}$Instituto de Nanociencia y Materiales de Arag\'on (INMA), CSIC-Universidad de Zaragoza, E-50009 Zaragoza, Spain
\linebreak
$^{g}$Departamento de F\'isica de la Materia Condensada, Universidad de Zaragoza, E-50009 Zaragoza, Spain
\linebreak
$^*$Corresponding authors: carlo.casari@polimi.it, jorge.lobo@csic.es, simona.achilli@unimi.it
\linebreak
$^\dag$These authors contributed equally
\end{flushleft} }

\twocolumn[
\begin{@twocolumnfalse}
	\maketitle
	\begin{abstract}
\noindent Graphdiyne-based carbon systems generate intriguing layered \textsl{sp}$-$\textsl{sp}$^2$ organometallic lattices, characterized by flexible acetylenic groups connecting planar carbon units through metal centers. At their thinnest limit, they can result in two-dimensional (2D) organometallic networks exhibiting unique quantum properties and even confining the surface states of the substrate, which is of great importance for fundamental studies. In this work, we present the on-surface synthesis of a highly crystalline 2D organometallic network grown on Ag(111). The electronic structure of this mixed honeycomb-kagome arrangement -- investigated by angle-resolved photoemission spectroscopy and scanning tunneling spectroscopy -- reveals a strong electronic conjugation within the network, leading to the formation of two intense electronic band-manifolds. In comparison to theoretical density functional theory calculations, we observe that these bands exhibit a well-defined orbital character that can be associated with distinct regions of the \textsl{sp}$-$\textsl{sp}$^2$ monomers. Moreover, we find that the halogen by-products resulting from the network formation locally affect the pore-confined states, causing a significant energy shift. This work contributes to the understanding of the growth and electronic structure of graphdiyne-like 2D networks, providing insights into the development of novel carbon materials beyond graphene with tailored properties.
	\end{abstract}
	\vspace{3mm}
	\vspace{5mm}
\end{@twocolumnfalse}
]

\begin{figure*} [t!]
	\centering
	\includegraphics[width=\textwidth,trim=2 6 4 2,clip]{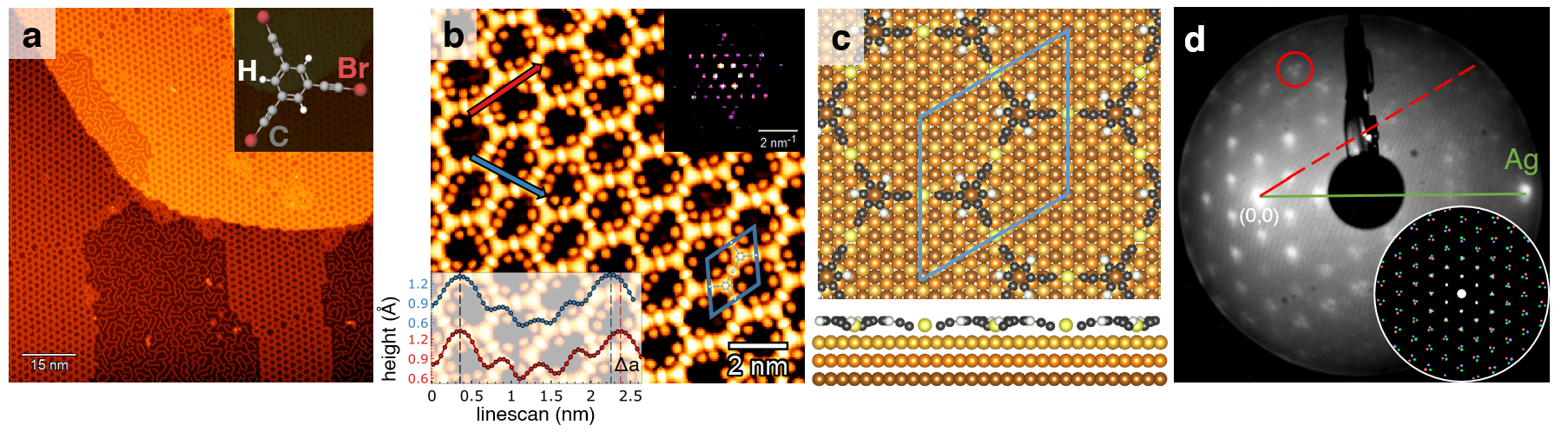}
	\caption{Structural characterization of 2D organometallic network on Ag(111). (a) Large-scale LT-STM image of the 2D network islands after annealing to ${\sim}$370~K, exhibiting long-range order for hundreds of nm$^2$ (STM set-point: $-$1~V, 50~pA). Inset: atomic model of the tBEB molecular precursor. (b) High-resolution LT-STM image of the mixed honeycomb-kagome network showing the linking Ag adatoms as the brightest spots, the monomers as triangular features connecting three Ag adatoms, and the Br atoms as blobs within the pores (STM set-point: 50~mV, 90~pA). The unit cell of the organometallic network (marked with a blue rhombus) contains two debrominated monomeric units and three Ag adatoms. Insets: 2D-FFT (top right); averaged line profiles along the two directions marked by the red and blue arrows (bottom left). (c) Simulated atomic structure of the organometallic network on Ag(111) in top and side views. Carbon atoms are shown in gray, hydrogen atoms in white, Ag adatoms in yellow, and substrate Ag in light-to-dark brown. (d) LEED pattern obtained at 28.6~eV, consisting primarily of a ${(4\sqrt{3}{\times}4\sqrt{3})}$R30$^{\circ}$ supercell over the Ag(111) lattice with a ${2.002\pm0.10}$~nm lattice constant. The red circle points to the pattern splitting into triplets (see text for details). Inset: incommensurate LEED pattern simulated by LEEDpat using the superlattice matrix \(\bigl( \begin{smallmatrix}+4.0 & -4.0\\ +3.6 & +7.6\end{smallmatrix}\bigr)\).}
	\label{structure}
\end{figure*}

\noindent Two-dimensional (2D) organometallic networks and covalent organic frameworks at their thinnest limit can be classified as quantum materials with relevant properties, such as the emergence of flat bands,\cite{Springer2020} topologically protected states,\cite{Wang:2013ag} superconductivity,\cite{Zhang2017} or 2D ferromagnetism.\cite{lobo2023} Such properties emerge in the presence of extended electronic conjugation of their building blocks.\cite{Liu2023,Hernandez-Lopez2021,Frezza2023,Galeotti2020, Vasseur2016,piquero2023} Among 2D systems, Graphyne and Graphdiynes stand out as novel 2D carbon materials beyond graphene, given their combination of \textsl{sp}- and \textsl{sp}$^2$-hybridized carbon atoms.\cite{Li2023} This leads to the emergence of diverse complex structures with distinct electronic properties, ranging from metals to tunable bandgap semiconductors and semi-metals. Their extended $\pi$-electron conjugation and topology produce multiple Dirac cones and band-manifolds that are key to their intriguing properties.\cite{Malko2012, Serafini2021JPCC} 

Despite several outstanding theoretical predictions,\cite{li2014graphdiyne,ge2018review} these systems were initially regarded as elusive and impractical. Such belief was proved wrong as the synthesis of multilayer graphdiyne was experimentally achieved.\cite{Li2010} This material presented remarkable properties and held promise for many advanced applications.\cite{chen20171d,jia2017synthesis, gao2019graphdiyne,zuo2019synthesis} Later on, the synthesis of 2D single-layer graphdiyne was also realized through on-surface synthesis (OSS).\cite{sun2016dehalogenative} The OSS technique relates to ``solvent-free'' reactions exploiting the catalytic activity of a metal surface to promote the formation of new bonds from smaller precursors.\cite{Lackinger2017,Clair2019} In the case of single-layer graphdiyne-like structures, the molecular precursors contain aromatic \textsl{sp}$^2$ carbon groups with linear acetylenic units that are usually terminated by a halogen atom. The de-halogenation of these precursors and the formation of an organometallic network is promoted by the native atoms from the metal surface.\cite{yang2020metalated,rabia2020structural, shu2020atomic} Subsequent thermal treatment can at times induce de-metalation and homocoupling with the formation of an all-carbon structure.\cite{sun2016dehalogenative,yang2020metalated, klappenberger2015surface, Klappenberger2018,Rabia-Nanoscale12019,Sedona-PCCP2020,rabia2020structural}

Although OSS is widely used in the synthesis of graphene-based nanoarchitectures,\cite{Wang2019} the fabrication of extended graphdiynes is still very challenging.\cite{sun2016dehalogenative,rabia2020structural} Nevertheless, achieving extended 2D structures with long-range order is critical to exploring the electronic structure of these layers beyond the generalized use of local electronic techniques. Indeed, the band structures of 2D graphdiyne-based organometallic networks are nowadays only accessible through calculations.\cite{yang2020metalated, Serafini2021JPCC,achilli2021graphdiynes, Serafini2021} They show the presence of electron conjugation throughout the organometallic layer and indicate that the metal centers induce hybridization and charge transfer to the network. Thus, it is pressing to experimentally validate these predictions, which are at the root of the emergence of multiple Dirac-cone band-manifolds that are key to many quantum properties of interest. Moreover, the doping role of the cleaved Br atoms existing on the surface as reaction by-products must be revisited,\cite{Piquero2018,yang2020metalated} as these could extend their influence to the confined surface state (SS) electrons at the pores\cite{PiqueroRMP2022} -- \textit{i.e.} the hexagonal holes contoured by the network exposing the underlying Ag surface.

In this study, we answer these intriguing fundamental questions. We present the structural and electronic characterization of an on-surface synthesized 2D graphdiyne-like organometallic network on Ag(111). The structure is characterized by scanning tunneling microscopy (STM) and low-energy electron diffraction (LEED), demonstrating the ability to achieve extended, long-range ordered structures. This allows us to experimentally explore the occupied electronic structure by means of angle-resolved photoemission spectroscopy (ARPES) and compare it to the site-dependent local density of states (LDOS) obtained from scanning tunneling spectroscopy (STS). Direct comparison with theoretical \textit{ab~initio} calculations by density functional theory (DFT) allows us to confirm the presence of extended $\pi$-electron conjugation and also deduce the dominant orbital character involved in each of these experimental bands. Finally, we show that the cleaved Br atoms generated during the OSS process not only electronically dope the 2D network, but also produce a local energy shift in the pore-confined surface state.

\section*{Results}

\textbf{Network formation and morphology.} Our molecular precursor of choice is 1,3,5-tris(bromoethynyl)benzene (tBEB), consisting of a benzene ring functionalized with three non-adjacent acetylenic groups ending with a bromine atom (see inset of Fig.~\ref{structure}(a)). This molecule has been used to produce 2D structures \textit{via} on-surface synthesis (OSS) on both Au(111) and Ag(111).\cite{sun2016dehalogenative,rabia2020structural, yang2020metalated,shu2020atomic} When deposited at room temperature (RT) on Ag(111), the debromination occurs spontaneously and the Br atoms are replaced by native Ag atoms from the substrate that bind to the alkynyl groups. The dominant structure formed is a 2D organometallic network coordinated by Ag adatoms, which connect two adjacent benzene units through their acetylenic groups and form $-$C$\equiv$C$-$Ag$-$C$\equiv$C$-$ bonds. This well-organized, self-assembled \textsl{sp}$-$\textsl{sp}$^2$ 2D network leads to hexagonal pores enclosing a different number of detached Br atoms, as shown in the close-up STM image of Fig.~\ref{structure}(b). Statistically, there is an average of 5.1 Br atoms per unit cell or, equivalently, per pore (Fig.~S1(b)). The missing 0.9 Br atoms from the six released during the network formation aggregate on the molecule-free areas of the Ag(111) surface in wiggly 1D arrangements (Fig.~S1(d)). 

The 2D network structurally forms a mixed honeycomb-kagome network, where the debrominated tBEB molecules (monomers) occupy the vertices of the pore contour, while the Ag adatoms (brighter spots in Fig.~\ref{structure}(c)) are in the middle of its sides. In this configuration, the organometallic bond is considered practically covalent in strength.\cite{yang2018two} Indeed, we find this 2D network to be stable at RT, according to our STM characterization at ${\sim}$300~K (Fig.~S1(e,f)). Annealing to ${\sim}$370~K results in the improved crystallization of the 2D network in terms of larger domain size and increased coverage (Fig.~\ref{structure}(a) and S1(a)), leading to extended long-range order as extracted from the well-defined LEED pattern shown in Fig.~\ref{structure}(d). Increasing the post-annealing temperature to ${\sim}$420~K results in a decrease in overlayer coverage, which we interpret as desorption of the monomeric units. Differently from the OSS on Au(111),\cite{sun2016dehalogenative,rabia2020structural} increasing further this temperature (up to 450--470~K) does not promote the network demetalation and the C--C homocoupling among the acetylenic units. Instead, at such higher temperatures we observe the disruption of the network and the formation of non-planar, globular islands (Fig.~S1(h)).

We can extract the average parameters of the 2D network from the 2D fast Fourier transform (2D-FFT) (inset of Fig.~\ref{structure}(b)), resulting in a mean periodicity of ${1.953~{\pm}~0.15}$~nm and unit vector angles of ${60.5^\circ~{\pm}~1.5\degree}$. These values are in agreement with previous works.\cite{yang2020metalated, sun2016dehalogenative,rabia2020structural,shu2020atomic} However, the flexibility of the alkynyl groups and the side interaction with a variable number of cleaved bromine atoms within the pores lead to slight directional distortions that affect the overall lattice periodicity and introduce some degree of structural anisotropy. In particular, the line profiles in the inset of Fig.~\ref{structure}(b) reveal a difference of ${\Delta a=0.12}$~nm. Such deviation is also visible in LEED patterns (Fig.~\ref{structure}(d)). In a first approximation, the overall LEED arrangement is hexagonal, exhibiting a ${(4\sqrt{3}{\times}4\sqrt{3})}$R30$^{\circ}$ supercell over the Ag(111) surface and yielding a lattice constant of 2.002~nm. However, the LEED pattern shows split higher-order diffraction spots that are more evident as we move away from the (0,0) spot due to the presence of rotationally non-equivalent domains. We obtain an excellent match of this distorted hexagonal superstructure when using nonequivalent unit vectors of 2.010~nm and 1.895~nm -- as extracted from the line profiles of the Fig.~\ref{structure}(b) inset -- in the simulated pattern (LEEDpat software,\cite{leedpat} see inset of Fig.~\ref{structure}(d)).

\begin{figure*} [h!]
	\centering
	\includegraphics[width=.85\textwidth,trim=2 6 2 2,clip]{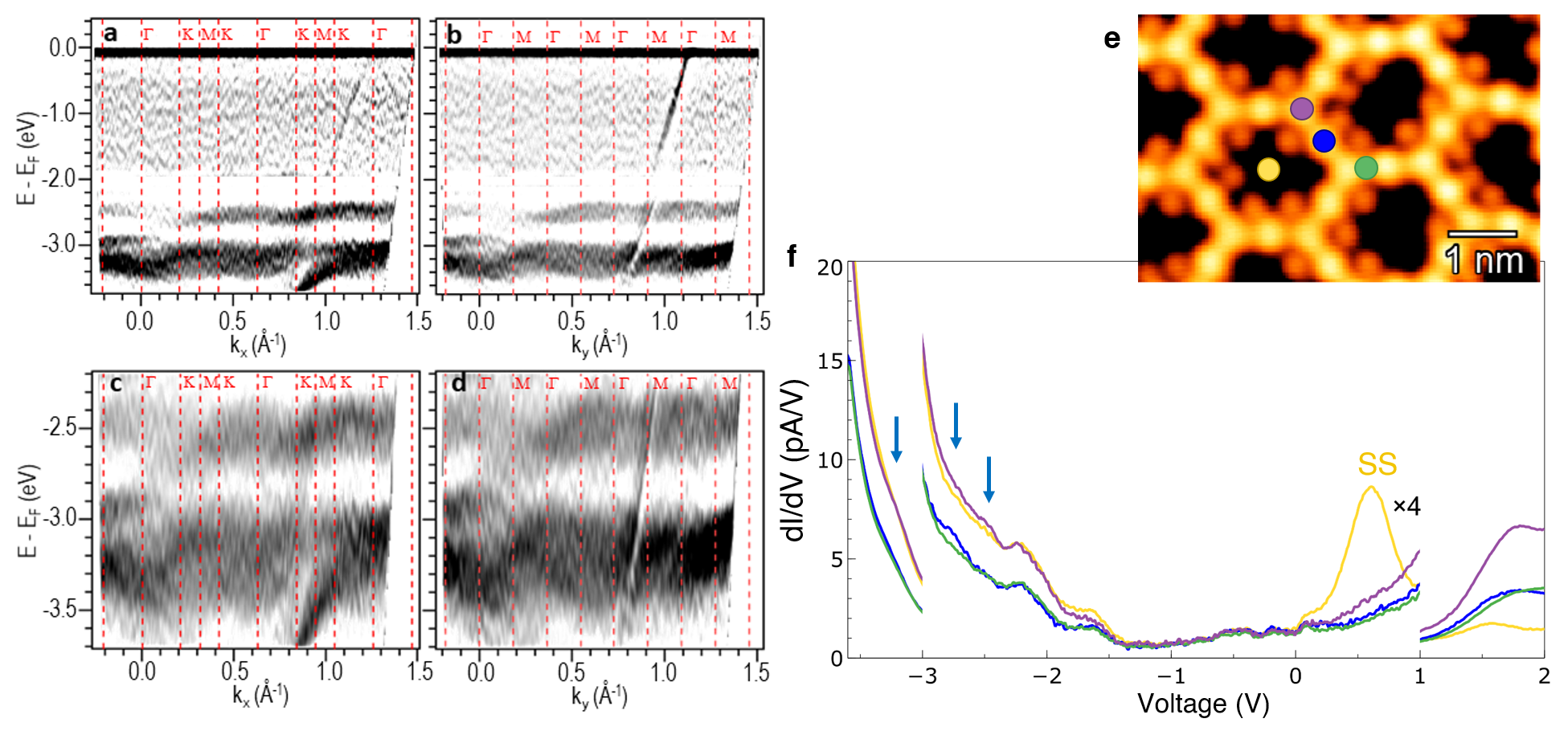}
	\caption{Electronic characterization of the 2D organometallic network on Ag(111). (a,b) Occupied electronic structure of the 2D network close to saturation on Ag(111) after annealing to ${\sim}$370~K. The second derivative of the ARPES spectral function was recorded along the (a) $\Gamma$K ($k_y = 0$) and (b) $\Gamma$M ($k_x = 0$) directions of the network, using a photon energy of 21~eV and keeping the sample at 120~K. We observe two dominant band-manifolds with relatively flat appearance related to the 2D network in the $-$2.25 to $-$3.5~eV region, as well as highly dispersive bands at large $k_x$ and $k_y$ values originating from the \textsl{sp} bands of silver (Fig.~S3). (c,d) Blown-up panels at the network's band-manifolds exhibiting dispersion. In panels (a--d), the color grayscale is linear (the darker, the more intense) and the vertical dashed lines indicate high-symmetry points of the 2D network. Note that the high-symmetry directions of the substrate are rotated by 30$^\circ$. (e) LT-STM image of the STS grid area (STM set-point: $-$100~mV, 50~pA). (f) Averaged LT-STS curves (from 10--15 spectra) taken over several equivalent points of the 2D network, as indicated by the color-coded dots in (e). The STS curve of the benzene rings is shown in purple, Ag adatoms in blue, acetylenic triple bonds in green, and pore centers in yellow. Note that the central region (from $-$3.0 to $+$1.0~V) has been multiplied by a factor of 4 for improved visualization. The blue arrows identify peaks within the ARPES band-manifold range, while "SS" identifies the pore-confined Shockley peak (STS set-point: $-$500~mV, 14~pA; ${V_{\text{RMS}}=15.5}$~mV and ${f_{\text{osc}}=817.3}$~Hz).}
	\label{ARPES}
\end{figure*}

We can obtain further structural information on the hexagonal network by DFT calculations for the 2D network in the gas phase and on top of a ${(4\sqrt{3}{\times}4\sqrt{3})}$R30$^{\circ}$ supercell of the Ag(111) substrate. In the gas phase, the calculated free-standing lattice constant of the network is 2.090~nm. In order to match the network with the underlying substrate (whose theoretical equilibrium lattice constant is ${a_{Ag}=0.295}$~nm), the network is slightly compressed to a lattice constant of 2.042~nm, in agreement with our STM results (within the experimental error). In virtue of this, the on-surface synthesized 2D network experiences a ${\sim}$2.2\% stress with respect to its gas phase equilibrium structure. Additionally, the interaction between the Ag adatoms and the substrate leads to bent monomers, as the central phenyl groups protrude from the surface while the linear acetylenic groups point towards it, as shown in Fig.~\ref{structure}(c). The overall corrugation of the network amounts to 0.05~nm, being defined as the height difference between the phenyl rings and the carbon atom in the acetylene moiety adjacent to the Ag adatom. Experimentally, the apparent corrugation as measured by LT-STM is 0.035~nm (with a set-point of $-$3.5~V~/~800~pA, see Fig.~S2(b)).

\begin{figure*} [t!]
	\centering
	\includegraphics[width=.9\textwidth,trim=6 10 0 2,clip]{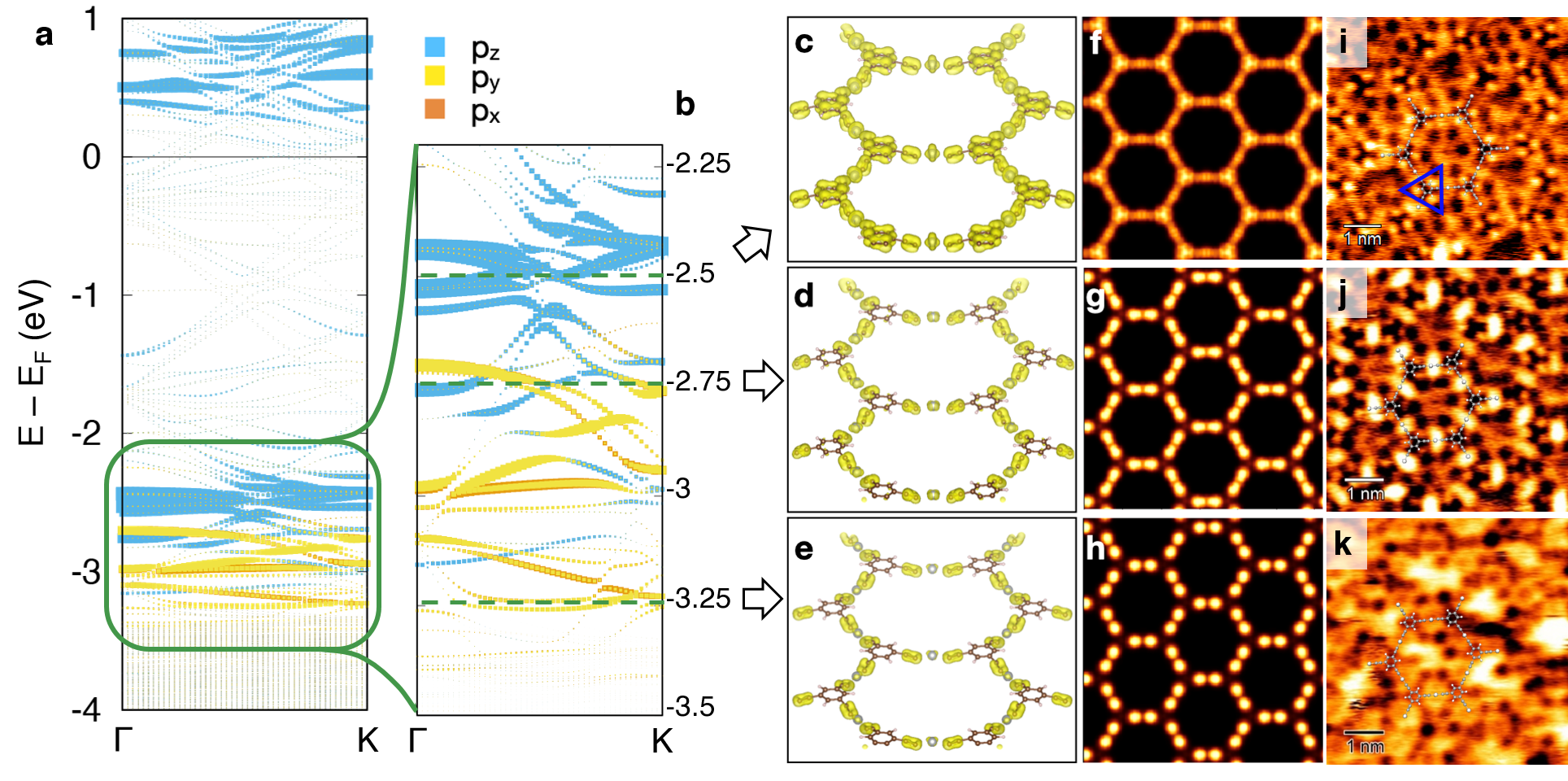}
	\caption{Comparison of the theoretical and local experimental electronic properties of the 2D organometallic network on Ag(111). (a,b) Band dispersion along a high-symmetry path in the network's Brillouin zone, showing resolved contributions for \textsl{p$_z$} (blue) and \textsl{p$_{x,y}$} (orange, yellow) states. (b) Magnification of (a) in the $-$2.2 to $-$3.5~eV range. (c$-$e) Isovalue maps of the calculated LDOS of the organometallic network on Ag(111) integrated in an energy window of 0.15~eV and centered at (c) $-$2.50~eV, (d) $-$2.75~eV, and (e) $-$3.25~eV (Ag substrate not shown). Isosurface level: 0.002~states~eV$^{-1}$~\AA$^{-3}$. (f$-$h) Simulated \(dI/dV\) maps at the same energies of panels (c--e) and at a constant height of 2~nm from the plane of the benzene rings. (i--k) Constant-current \(dI/dV\) maps acquired at 4.8~K and the same bias as panels (c--e), with ${V_{\text{RMS}}=10.0}$~mV and ${f_{\text{osc}}=817.3}$~Hz (STS set-point: (i) $-$2.50~V, 350~pA; (j) $-$2.75~V, 400~pA; (k) $-$3.25~V, 450~pA). The dominant intensity is associated with the phenyl rings at $-$2.50~V (see the orbital arrangement in the blue triangle of (i)), with the metal centers at $-$2.75~V, and with the acetylenic units at $-$3.25~V.}
	\label{electronic}
\end{figure*}

\noindent\textbf{Electronic characterization by ARPES.} The excellent growth quality of the network allowed us to determine the occupied electronic structure employing an averaging technique, such as ARPES (Figs.~\ref{ARPES}, S3 and S4). Meaningful measurements at the BaDElPh beamline\cite{petaccia2009bad} of the Elettra Synchrotron were obtained after achieving a sample preparation close to a full monolayer coverage of the mixed honeycomb-kagome network on the Ag(111) surface, which was followed by thermal annealing to ${\sim}$370~K. From a visual inspection of Fig.~\ref{ARPES}, we can easily identify the presence of two shallow-dispersive band-manifolds in the energy range between $-$2.25~eV and $-$3.5~eV, and highly dispersive features that reach the Fermi level (E\textsubscript{F}) at high parallel momenta (darker grey corresponding to higher photoemission intensity). A comparison with the pristine substrate (reported on the right side of Fig.~S3) evidences that these highly dispersive features are related to the bulk Ag \textsl{sp}-band. Notably, we find that the Ag(111) surface state at ${k=0}$, clearly appearing in the pristine substrate, is not visible anymore, confirming its electronic depopulation upon network formation.\cite{yang2020metalated}

The two-dimensional character of the band-manifold (Fig.~\ref{ARPES}(c,d)) is deduced by their invariability after performing a photon energy dependence check (Fig.~S4). Based on previous work,\cite{Hernandez-Lopez2021} these shallow-dispersive bands are assigned to conjugated kagome multi-bands arising from honeycomb networks. Compared to free-standing calculations (Fig.~S9), in which the band manifold is around the Fermi level, this kagome multi-band is considerably downshifted (by more than 2~eV) by the Ag(111) substrate. Although umklapps of the \textsl{sp}-bulk bands are absent, their quality appears to be significantly better than what has been reported to date,\cite{Hernandez-Lopez2021} to the point that it is visible even in the direct ARPES signal (Figs.~S3 and S4). These intense kagome multi-bands are likely related to the very strong organometallic bond existing between the molecules and the Ag adatoms.\cite{yang2020metalated} Contrarily, the umklapps absence relates to the observed distortions of this network compared to other strictly periodic organometallic systems.\cite{Hernandez-Lopez2021}

\noindent\textbf{Assessment of the electronic structure by DFT.} To corroborate the existence of this kagome multi-band we performed DFT calculations of the 2D network on top of the Ag(111) substrate. The band structure along a high symmetry direction of the ${(4\sqrt{3}{\times}4\sqrt{3})}$R30$^{\circ}$ supercell is reported in Fig.~\ref{electronic}(a). Superimposed to the bands of the whole system, consisting of the 2D network and the substrate (dotted lines), the \textsl{p}-bands of carbon stand out, plotted with different colors depending on their orbital symmetry. The calculation confirms the presence of electronic bands in ${\sim}$1~eV-wide energy intervals above the top of the Ag \textsl{d}-bands and above the Fermi level. The overall shift of more than 2~eV observed between the experiment and the calculated free-standing band structure (Fig.~S9) is theoretically reproduced in the simulation of the Ag(111)-supported organometallic network.

Fig.~\ref{electronic}(b), showing an enlargement of the occupied states region, highlights two groups of energy bands. The first, at larger binding energy, exhibits a \textsl{p$_{x,y}$} character and can be identified with the band-manifolds centered at $-3.25$~eV in ARPES. The second, with smaller binding energy, displays a \textsl{p$_z$} character and corresponds to the flat band-manifolds experimentally centered around $-2.5$~eV (Fig.~\ref{ARPES}). This assignment is possible due to the reasonably good energy correspondence between experimental and calculated bands, and the slightly different dispersion of the two sets of bands. As in the experiments, we find that the \textsl{p$_z$} states (top band) are flatter, while the \textsl{p$_{x,y}$} states are characterized by more dispersive features around the $\Gamma$ point. Significantly, both the expected Dirac-cone features (typical of kagome networks, observed in the freestanding band structure in Fig.~S9) and the spin polarization of the states are practically destroyed. This is caused by the interaction with the substrate, which is more pronounced for \textsl{p$_z$} states than \textsl{p$_{x,y}$} states due to their different extension in the vertical direction. Note that the theoretical band structure is folded into the irreducible $\Gamma$K path of the ${(4\sqrt{3}{\times}4\sqrt{3})}$R30$^{\circ}$ supercell, whereas ARPES data is displayed along two extended high-symmetry directions ($\Gamma$K and $\Gamma$M) of the network's Brillouin zone.

We also performed DFT calculations of the network with Br atoms within the pores. In these calculations, the halogens are strongly bound to the Ag surface in positions close to the benzene's hydrogens between acetylenic groups, \textit{i.e.} where they are most frequently observed at the STM (Fig.~S2(e--f)). Br-related bands appear in the energy interval between $-$3~eV and $-$2~eV, as shown in Fig.~S8(c). Despite being energetically very close to the carbon network's bands, they have little effect on the latter in terms of energy shift and dispersion, which is consistent with the hybridization of Br atoms mainly with the substrate's Ag atoms. Although bromine atoms have been reported to induce an upwards rigid shift of about 200~meV (similar to \textsl{n}-doping effect) in conjugated poly(para-phenylene) systems,\cite{basagni2016tunable,Piquero2018} our theoretical results do not show evidence of such an effect in the Ag(111)-supported 2D network, consistently with a negligible value of the computed charge transfer between Br and C atoms in the present case.

\begin{figure*} [h!]
	\centering
	\includegraphics[width=.7\textwidth,trim=6 12 6 2,clip]{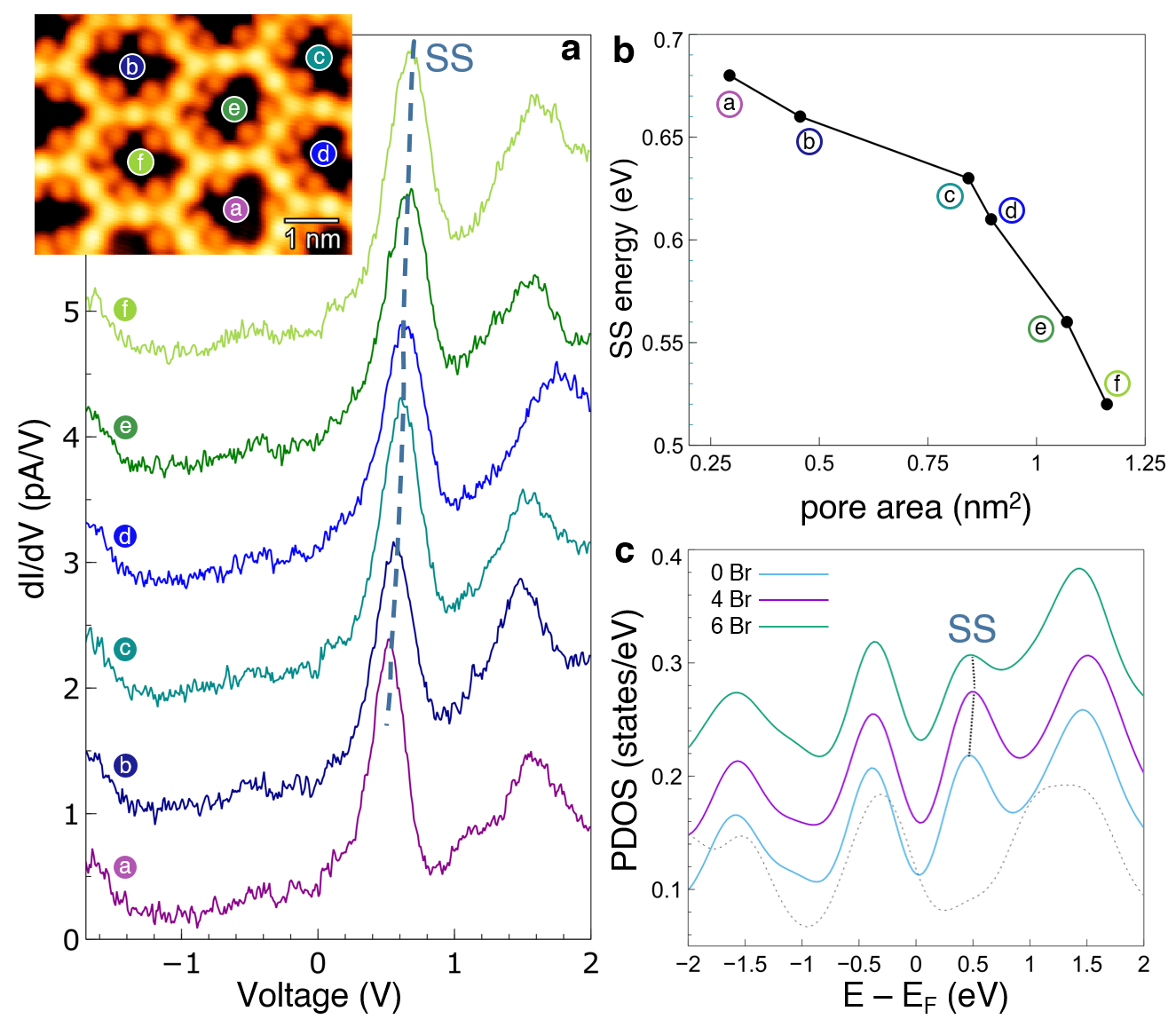}
	\caption{(a) Single-point LT-STS curves acquired at the center of pores containing a progressively higher number of Br atoms, as indicated by the color-coded dots in the inset (STS set-point: $-$500~mV, 14~pA; ${V_{\text{RMS}} =15.5}$~mV and ${f_{\text{osc}}=817.3}$~Hz). Inset: reference LT-STM image (STM set-point: $-$100~mV, 50~pA). (b) Correlation between the exposed pore area subtracting the Br atoms (labeled as the pores in the inset of (a)) and the maximum SS peak energy. (c) Simulated PDOS at the center of the pores (surface Ag atoms) with increasing Br atom occupancy exhibiting the experimental energy shift. The SS is absent in the dashed curve at the bottom, which refers to the PDOS on a Ag atom in the middle of the substrate slab.}
	\label{poreSS}
\end{figure*}

\noindent\textbf{Electronic characterization by LT-STS.} To corroborate and expand the experimental results on the electronic features obtained by ARPES, we acquired differential conductivity ($dI/dV$) maps and constant-height (C.H.) images through LT-STM/STS at 4.8~K. The complete datasets, reported in the S.I. (Figs.~S5 to S7), include sample mapping in the $-$3.50 to 2.00~V range at both constant current and constant height, as well as a collection of energy slices of a LT-STS map obtained from a 36${\times}$24-pixel grid. The corresponding LT-STM image of the $6{\times}4$~nm$^2$ scanned area and selected point spectra are reported in Fig.~\ref{ARPES}(e,f). Each curve of the $dI/dV$ spectra is averaged over 10--15 equivalent points of the network.

Notably, we reproduce the LT-STS data together with the results obtained by ARPES below the Fermi level (E\textsubscript{F}). Particularly, we find weak peaks and shoulders at $-$3.25~V, $-$2.75 and $-$2.50~V, marked as blue arrows in Fig.~\ref{ARPES}(f). Excluding the peak at $-$2.75~V, which is localized around the Ag adatoms, the other two peaks would have been overlooked without the ARPES datasets. The corresponding constant-current $dI/dV$ maps, obtained from the LDOS at these relevant energies, are shown in Fig.~\ref{electronic}(i--k). Evidently, different orbitals are accessed at each bias, indicating the existence of distinct electronic states, as will be discussed in the following. Note that the electronic features visible approximately at $-$2.25~V and $-$1.60~V are associated with tip states, since their shape and energy remain unvaried at all sites in the STS grid, but are absent in ARPES. Moving to the empty electronic region (positive bias) we find energy states as identified in a previous work;\cite{yang2020metalated} in particular, we observe the confined surface state (SS) at 0.60~V, and a state localized around the benzene rings at 1.80~V. A more in-depth analysis follows in the Discussion section.

\section*{Discussion}
The 2D organometallic network has a strong semiconductive character. The calculation for the free-standing network reveals that the unsupported network would be a half-metal with a calculated energy gap for the majority spin component of ${\sim}$3~eV (Fig.~S9(a)). The frontier orbitals have mixed \textsl{p$_{x,y}$}--\textsl{p$_z$} character in the occupied region of the energy spectrum, and \textsl{p$_z$} character in the unoccupied one. A similar energy separation is found as a pseudo-gap in the empty region of minority spin bands, which are shifted upward by about 0.5~eV with respect to the majority manifold. When the 2D network is formed on Ag(111), the spin imbalance is lost as a consequence of the interaction with the substrate, and the network bands undergo a shift to lower binding energies. Such shift relates to a net charge transfer of 0.5~eV/unit cell towards the network, resulting from an increase of charge on the C and H atoms (0.8~eV/unit cell) and a depletion of 0.36~eV/unit cell from the Ag adatoms. The experimental estimate of a 3.10~eV energy gap can be deduced from ARPES data, showing no electronic states down to approximately $-$2.25~eV (Fig.~\ref{ARPES}(a--d)), and from the onset of the first unoccupied network state, starting at ${\sim}$0.85~eV according to STS (Fig.~\ref{ARPES}(f)). Note that we discard the states related to the confined pore states (peaking at 0.6~eV) as they are substrate-dependent. Indeed, the $dI/dV$ maps of Fig.~S5(g--o) show a reduced electronic intensity at the network from $-$2.25~eV to 0.6~eV, confirming this large energy gap. Likewise, DFT calculations for the Ag(111)-supported network (Fig.~\ref{electronic}(a)) show an absence of relevant states on the 2D network between $-$2.25 and 0.3~eV, corresponding to an energy gap of 2.55~eV. Considering the expected gap underestimation for DFT calculations,\cite{GW2007} this calculated value agrees with the experiment.

In the following, we focus our attention on the occupied low-energy states at the energy position of the ARPES bands and compare them with the experimental STS (Fig.~\ref{ARPES}(f)) and $dI/dV$ maps (Fig.~\ref{electronic}(i--k)). We also compare these experimental results with the simulated $dI/dV$ maps (f--h) and the calculated LDOS isosurfaces cut at a specific height (c--e), showing the states' spatial symmetry. At $-$2.50~V bias, we detect a weak shoulder on the STS at the benzene ring position (purple curve in Fig.~\ref{ARPES}(f)). The dominant contribution of the $dI/dV$ map at this energy (Fig.~\ref{electronic}(i)) corresponds to three circular features in a triangular arrangement located at the carbon atoms covalently bound to hydrogens (between the acetylenic groups) in the benzene rings -- see the blue triangle in Fig.~\ref{electronic}(i). Note that these features are strongly modulated in intensity by the presence of neighboring cleaved Br atoms. The simulated LDOS at this energy (Fig.~\ref{electronic}(f)) displays such molecular orbital (although with differences in the intensity distribution due to the inaccurate comparison of constant-current conditions with constant-height calculations). This orbital, according to the isosurfaces in Fig.~\ref{electronic}(c), exhibits mainly a \textsl{p$_z$} character.

Moving down in energy, at $-$2.75~V we observe the main peak at the Ag adatom position of the STS spectra (blue curve in Fig.~\ref{ARPES}(f)), which is clearly localized at the same position in the $dI/dV$ map (Fig.~\ref{electronic}(j)). The surrounding Br atoms do not seem to interfere with this state, which is also evident in C.H. maps (Fig.~S6(d)). The simulated LDOS (Fig.~\ref{electronic}(d,g)) localizes the electronic states at the acetylenic groups and in the proximity of Ag adatoms, thus revealing that the organometallic linking contributes at this energy to the overall electronic structure. This is confirmed by the projected density of states (PDOS), evaluated at the $\Gamma$ point (Fig.~S8(a)), \textit{i.e.} for slowly decaying states, showing the energy coincidence of states on triple bonds and Ag adatoms. Here, the orbital character is a mixture of \textsl{p$_z$} and \textsl{p$_y$} states for C, and of \textsl{s} and in-plane \textsl{d} states for Ag. We attribute the enhanced intensity of the adatom to the strong hybridization with the substrate and to its expected outer \textsl{s}-character, which would sharply increase the tunneling probability.

At $-$3.25~eV, the $dI/dV$ map exhibits most of its periodic signal as an extended halo centered at the Ag adatoms, amplified at the position of the triple C$\equiv$C bonds (Fig.~\ref{electronic}(k)). Indeed, we find a very subtle shoulder at the acetylenic group (green curve in Fig.~\ref{ARPES}(f)) of the STS. This feature is much more evident in the energy slice at $-$3.25~eV extracted from the STS grid in Fig.~S7(b), from which we deduce it is entirely an electronic state of the 2D network. Note that observing by STS such a subtle peak in the occupied region, so far away from the zero bias and so close to the substrate's \textsl{d}-bands, is an impressive experimental challenge. We find excellent agreement with the LDOS calculations and isosurfaces (Fig.~\ref{electronic}(e,h)), showing that the charge is located at the triple bonds with a mixture of \textsl{p$_x$} and \textsl{p$_y$} orbital character.

This comparison between experiment and theory allows not only to identify the orbital character of the bands, but also to spatially define the dominant contribution within the 2D network. In particular, the band-manifold closest to E\textsubscript{F} (around $-$2.50~eV) resonates at the central benzene of the monomers with an out-of-plane (\textsl{p$_z$}) orbital character, whereas the deeper band-manifold (around $-$3.25~eV) predominantly activates the acetylenic groups coordinated to the metal centers in a mixture of in-plane (\textsl{p$_x$} and \textsl{p$_y$}) orbitals.

Coming to the unoccupied states, we find electronic contributions related to the 2D network for voltages in the 1.60 to 2.00~V range (Fig.~\ref{ARPES}(f)). At these energies, the intensity is mostly localized around the benzene rings (Fig.~S5(s--u)), similarly to what was observed at $-$2.50~V (Fig.~\ref{electronic}(i)). Nevertheless, a progressive change in the orbital character of these states is observed as the energy increases. The $dI/dV$ signal changes from homogeneously triangular at the carbon atoms linked to the hydrogens, to a smeared, almost circular shape. In contrast, the intensity of the Ag adatoms gradually decreases, as also confirmed by C.H. imaging (Fig.~S6(j,k)). Indeed, at 2.00~V the dominant intensity is fully located on the benzene rings, with a \textsl{p$_z$} orbital character as demonstrated by the isosurfaces shown in Fig.~S10.

Finally, the prominent feature of the confined Shockley state peaking at 0.60~V (yellow curve in Fig.~\ref{ARPES}(f)) deserves further attention. The main energy shift is dependent on the pore size, shape, and barrier potential landscape created by the 2D network.\cite{muller2016confinement,PiqueroRMP2022} In this system, the network is able to completely deplete the Ag SS, as the ARPES cannot detect it below E\textsubscript{F}. On top of this, the Br atoms could also influence this shift, as previously reported.\cite{yang2020metalated} Indeed, we find that the number and position of bromine atoms are key to understanding the observed SS shift. Increasing the number of Br atoms moves the peak energy from 0.51 to 0.69~V (Fig.~\ref{poreSS}(a)), but we could find shifts as far as 0.85~V. Such energy differences are visible as intensity modulations in the $dI/dV$ maps and energy slices of the STS grid of Figs.~S5(n--p) and S7(g--i). Notably, we observe an inverse correlation between the SS energy and the pore area, which strictly depends on the Br atom pore occupancy (Fig.~\ref{poreSS}(b)). In this analysis, the pore area is measured as the region of exposed silver inside each pore after subtracting the Br atom space. Such SS shift due to the presence of Br atoms is also reproducible by DFT. The projected density of states (PDOS) for Ag of Fig.~\ref{poreSS}(c) confirms the shift in the SS position, but it is however similar for both the four Br and six Br situations. This is likely because the simulation does not take into account the distortion of the network caused by the randomly placed bromine atoms, which can significantly influence the real pore area.

\section*{Conclusions}
In summary, we show that we can synthesize a high-quality 2D organometallic network by evaporating tBEB molecules on Ag(111). By means of ARPES and STM/STS, we experimentally determine the electronic gap of this organometallic network, amounting to 3.10~eV. The observation of two evident band-manifolds manifests a high degree of electronic conjugation throughout the 2D network. DFT calculations shed light on these electronic features, confirming a non-negligible charge transfer from the substrate to the network. Notably, we identify the orbital character of each band-manifold and correlate it with their main spatial distribution, either out-of-plane at the benzenes or in-plane at the acetylenic groups. Finally, we elucidate the role of the residual Br atoms located within the pores and find that we can correlate the energy shift of the confined Shockley state with the available pore area (subtracting the space occupied by the halogens). These results further evidence that strong organometallic conjugation of these 2D networks is possible even on top of metallic surfaces. Moreover, they uncover the effect of the halogen byproduct of OSS reactions on the confined states of the underlying substrate. As materials based on \textsl{sp}-carbon show peculiar fundamental properties including topological states,\cite{topology-GDY-2020,Cirera-NatMat} our findings might be of great interest to future studies and could be key to the implementation of these networks in high-end applications.

\section*{Methods}
\textbf{Experiment.} All experiments were performed \textit{in situ} under ultra-high vacuum (UHV) conditions, with a base pressure in the $10^{-12}$ to $10^{-10}$~mbar range. Single-crystal Ag(111) (MaTeck) was employed for the on-surface synthesis (OSS) process unless otherwise specified. Low-temperature scanning tunneling microscopy and spectroscopy (LT-STM/STS) investigations were conducted with a Scienta Omicron microscope, cooled down to 4.8~K by liquid helium. LT-STS spectra and maps were obtained through a lock-in amplifier with an oscillation frequency of ${f_{\text{osc}}=817.3}$~Hz and a modulation amplitude of ${V_{\text{RMS}}=10{-}15.5}$~mV (root-mean-square), as indicated in figure captions. The lock-in output was compared to the numerical derivative of the $I(V)$ signal in order to express it in pA/V. Room temperature (RT) STM experiments were carried out with a variable-temperature microscope (VT-STM by Scienta Omicron), using homemade electrochemically etched tungsten tips and mica-supported polycrystalline Ag(111) substrates. Low-energy electron diffraction (LEED) and angle-resolved photoelectron spectroscopy (ARPES) measurements were carried out at 120~K at the BaDElPh beamline\cite{petaccia2009bad} of the Elettra Synchrotron in Trieste using an Omicron SpectaLEED and a SPECS Phoibos~150 hemispherical analyzer, respectively. In the ARPES experiments, the photon energy of the synchrotron radiation was varied in the 21--31~eV range; the total energy of the photoemitted electrons was set to 20~meV, while its angular resolution was below $0.3\degree$. STM data were analyzed using the image processing Gwyddion~2.63 software.\cite{nevcas2012gwyddion} The LEED pattern was simulated using the LEEDpat~4.3 software.\cite{leedpat}

The surfaces were prepared through several cleaning cycles of Ar$^{+}$ sputtering and annealing to 620$-$720~K until STM or LEED inspection revealed negligible traces of contaminants. The molecular precursor, \textit{i.e.} 1,3,5-tri(bromo\-ethynyl)benzene (tBEB), was synthesized as previously reported in the literature.\cite{mangione2013electrogenerated,liu2021transition} The precursor in powder form was loaded in a quartz crucible and inserted in an organic molecular evaporator (OME, provided by Dr. Eberl MBE-Komponenten). As the vacuum sublimation temperature is around RT, the crucible was kept at 303~K through a PID feedback control loop employing resistive heating and a water cooling system. After placing the substrate at RT in front of the OME opening, at a distance of about 10 to 20~cm, the OME shutter was kept open for several minutes (1 to 5~min, depending on setup), reaching a maximum chamber pressure of 8~$\times~10^{-8}$~mbar. The samples were subsequently annealed for 10--15~min to ${\sim}$370~K.

\noindent\textbf{Theory.} Theoretical calculations were performed by density functional theory (DFT) within the generalized gradient approximation (GGA) in the PBE form;\cite{PBE} van der Waals interactions between the organic overlayer and the substrate were included \textit{via} a DFT-D2 Grimme potential.\cite{Grimme} We used the approach implemented in the SIESTA code\cite{Sole02} that relies on norm-conserving pseudopotentials and an atomic-orbitals basis set. We adopted a double-zeta basis set with polarization orbitals and a mesh-cutoff of 400~Ry for the kinetic energy value that sets the real-space grid. The pseudopotential for Ag includes only \textsl{s} and \textsl{p} states in valence in order to achieve a better agreement with the experiments for the energy position of the \textsl{d} band.

The simulated STM images are obtained with the Tersoff-Hamann approach.\cite{Tersoff} The organometallic network matches a ${(4\sqrt{3}{\times}4\sqrt{3})}$R30$^{\circ}$ supercell of the underlying Ag(111) surface, as can be deduced from the STM images and the calculation of the free-standing network, which exhibits an equilibrium lattice constant equal to 2.090~nm; this leads to a 2.2\% compression when placed on Ag(111) (theoretical equilibrium lattice constant ${a_{Ag}=0.295}$~nm). We relaxed the organometallic layer and the first substrate layer until the forces reached the tolerance value of 0.04~eV~$\AA^{-1}$. We used a $3{\times}3$ sampling of the Brillouin zone. Along the $z$ direction, the slab includes three layers of the Ag substrate, the molecular overlayer, and ${\sim}85~\AA$ of vacuum. Upon relaxation, the slab thickness is increased to 9 layers of Ag for a better description of the electronic states.\\

\noindent{\bf Author contributions}\\
\noindent A.O.B. synthesized the molecular precursor. P.D., F.T., and J.L.-C. conducted the STM and LEED experiments, and P.D. analyzed the data. P.D., F.T., G.D.S., L.P., A.L.B., J.L.-C., and C.S.C. conducted the ARPES experiments, and J.L.-C. analyzed the data. S.A. conceived the theoretical modeling, performed the DFT calculations, and analyzed the data. G.O. contributed to the analysis of the theoretical results. P.D., S.A., J.L.-C., and C.S.C. wrote the manuscript. C.S.C. conceived the project. All authors contributed to the revision and final discussion of the manuscript.\\

\noindent{\bf Data Availability}\\
\noindent The data that support the findings of this study are available from the corresponding authors upon reasonable request.

\section*{Associated Content}

Supporting Information: additional LT- and RT-STM data (Fig.~S1-S2), including complete datasets of $dI/dV$ maps (Fig.~S5), constant height imaging (Fig.~S6), and STS grids (Fig.~S7); additional ARPES data showing the 2D network's signal contrasted with the Ag(111) substrate (Fig.~S3), and at different photon energies (Fig.~S4); additional DFT calculations simulating the atomic structure including cleaved Br atoms (Fig.~S2), the 2D network's density of states, band dispersion (Fig.~S8), spin band structure (Fig.~S9), and $dI/dV$ maps (Fig.~S10).

\section*{Acknowledgements}

The authors acknowledge Elettra Sincrotrone Trieste for providing access to its synchrotron radiation facilities and for financial support under the SUI internal project. P.D., F.T., A.L.B. and C.S.C. acknowledge funding by the European Research Council (ERC) under the European Union’s Horizon 2020 Research and Innovation Programme ERC-Consolidator Grant (ERC CoG 2016 EspLORE Grant Agreement No. 724610, website: \url{https:// www.esplore.polimi.it}). P.D., A.L.B. and C.S.C. also acknowledge the project funded under the National Recovery and Resilience Plan (NRRP), Mission 4 Component 2 Investment 1.3 – Call for tender No. 1561 of 11.10.2022 of Ministero dell’Universit\`a e della Ricerca (MUR); funded by the European Union – NextGenerationEU. Award Number: Project code PE0000021, Concession Decree No. 1561 of 11.10.2022 adopted by Ministero dell’Universit\`a e della Ricerca (MUR), CUP D43C22003090001, Project title ``Network 4 Energy Sustainable Transition – NEST''. S.A. and G.O. acknowledge CINECA for the use of supercomputing facilities under the IscraC program (project DECANET -- ID: HP10CN6MUU). S.A. and G.O. also thank the support of the project ``TIME2QUEST'' (Progetto di Iniziativa Specifica INFN). A.O.B. acknowledges support from the Ministero dell’Universit\`a e della Ricerca (MUR) and the University of Pavia through the program ``Dipartimenti di Eccellenza 2023--2027''. J.L.-C. acknowledges the use of Servicio General de Apoyo a la Investigaci\'on-SAI and the Laboratorio de Microscop\'ias Avanzadas of the Universidad de Zaragoza. J.L.-C. also acknowledges financial support from Grant References No. PID2019-107338RB-C64 and PID2022-138750NB-C21 funded by MCIN/AEI/10.13039/501100011033, by ``ERDF A way of making Europe'' and ``European Union NextGenerationEU/PRTR''. J.L.-C. also thanks the Aragonese Projects RASMIA E12\_20R co-funded by Fondo Social Europeo. \\

\noindent{\bf Competing interests}\\
\noindent The authors declare no competing financial interests.\\

\small
\bibliographystyle{ieeetr}
\bibliography{biblio}

\normalsize

\begin{figure*} [b!]
	\centering
	\includegraphics[width=.85\textwidth,trim=2 2 2 2,clip]{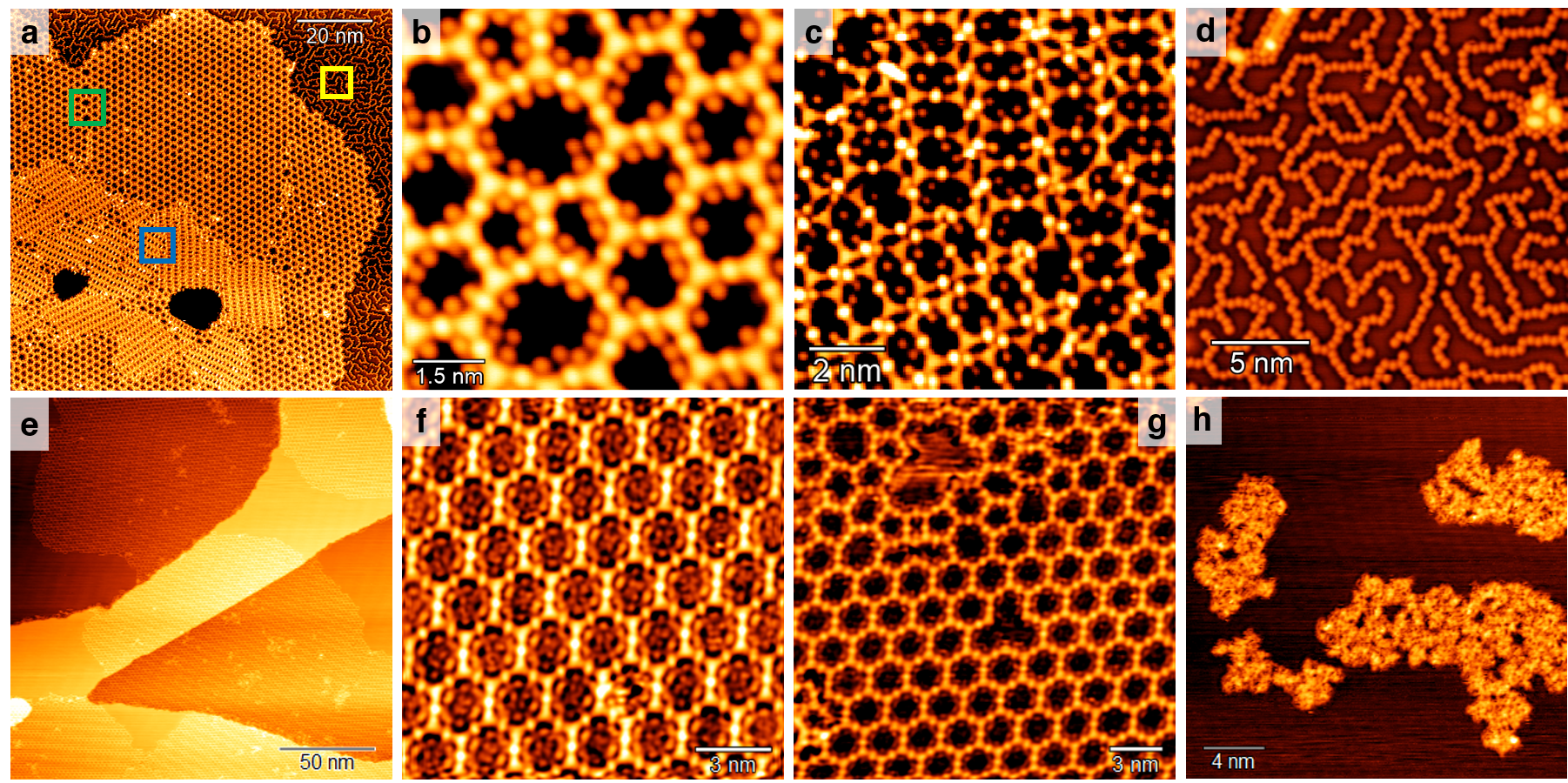}
	\caption*{Figure S1. Selection of LT and RT-STM images of 2D organometallic network on Ag(111), highlighting typical defects encountered alongside the mixed honeycomb-kagome 2D network. (a) Large-scale LT-STM image overview of the 2D network after annealing to ${\sim}$370~K, showing the coexistence of the porous and compact phase (STM set-point: 1.0~V, 50~pA). (b) Zoom-in at the green square of (a) showing a typical line defect at a grain boundary of coalescing 2D network islands consisting of aligned octagonal and pentagonal pores (STM set-point: $-$50~mV, 90~pA). (c) Zoom-in at the blue square of (a) displaying the residual compact phase (STM set-point: 70~mV, 100~pA). (d) Zoom-in at the yellow square of (a) showing the atomic arrangement of the cleaved Br atoms on the Ag(111) surface (STM set-point: $-$1.0~V, 50~pA). (e) Large-scale RT-STM overview image after molecular deposition and annealing on the Ag surface (STM set-point: 700~mV, 400~pA). (f) RT-STM image showing that the 2D network is stable at this temperature (STM set-point: $-$400~mV, 400~pA). (g) RT-STM image showing the emergence of two different defects with respect to the LT case: a single monomeric vacancy that does not disrupt the surrounding network, and a multiple monomeric vacancy aggregating into a larger network hole that distorts the adjacent pores into rectangles, pentagons, heptagons, or octagons to accommodate the deformation (STM set-point: 600~mV, 400~pA). (h) RT-STM image of the degraded network after annealing to ${\sim}$470~K (STM set-point: 700~mV, 400~pA).}
	\label{stm}
\end{figure*}

\begin{figure*} [b!]
	\centering
	\includegraphics[width=0.75\textwidth,trim=2 2 2 2,clip]{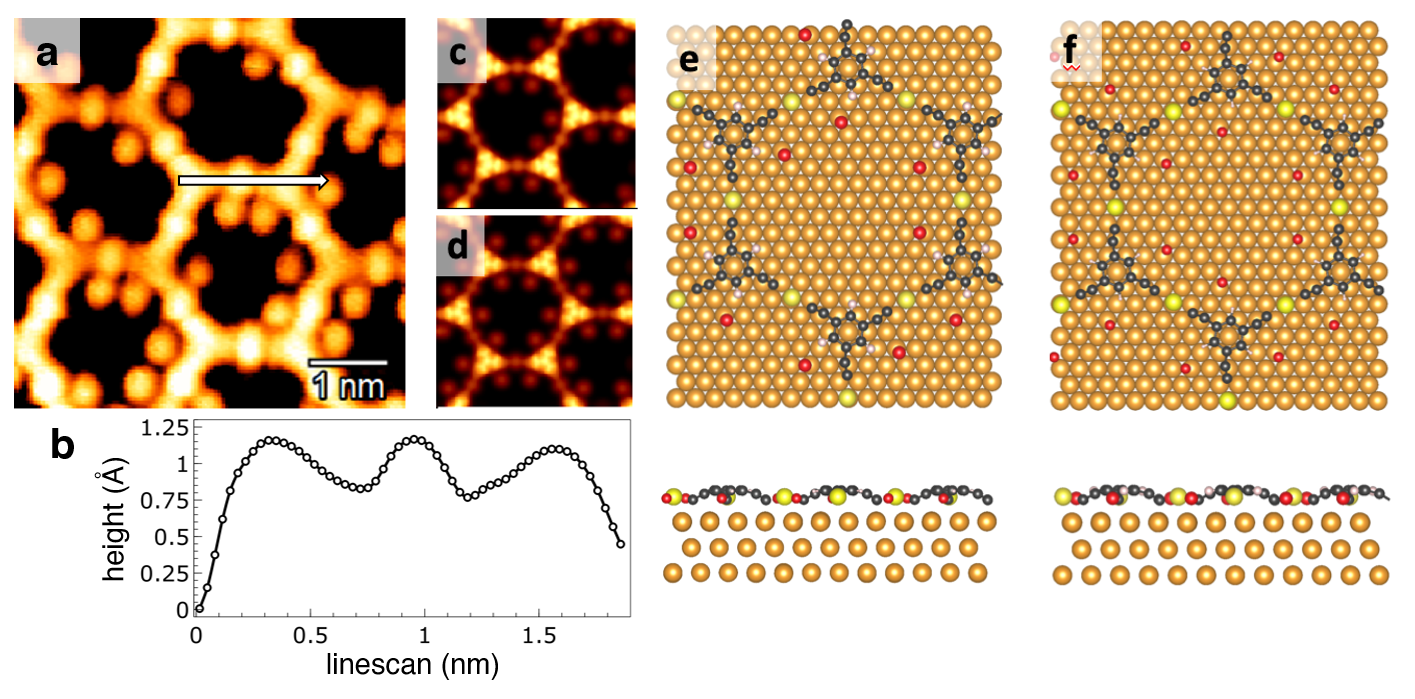}
	\caption*{Figure S2. (a) High-resolution LT-STM image (STM set-point: $-$3.5~V, 800~pA). (b) Average line profile measured along the direction indicated by the white arrow in (a). (c,d) Simulated STM images in the presence of four (c) and six (d) Br atoms in the network, obtained by integrating occupied states in an energy window 0.5~eV wide, at a distance of $2~\AA$ from the surface. (e,f) Simulated atomic structure of the Ag(111)-supported 2D organometallic network in top and side view with four (e) and six (f) Br atoms in each pore. Carbon atoms are in gray, Br atoms in red, Ag adatoms in yellow, and substrate Ag atoms in orange.}
	\label{detail}
\end{figure*}

\begin{figure*}
	\centering
	\includegraphics[width=0.85\textwidth,trim=2 2 2 2,clip]{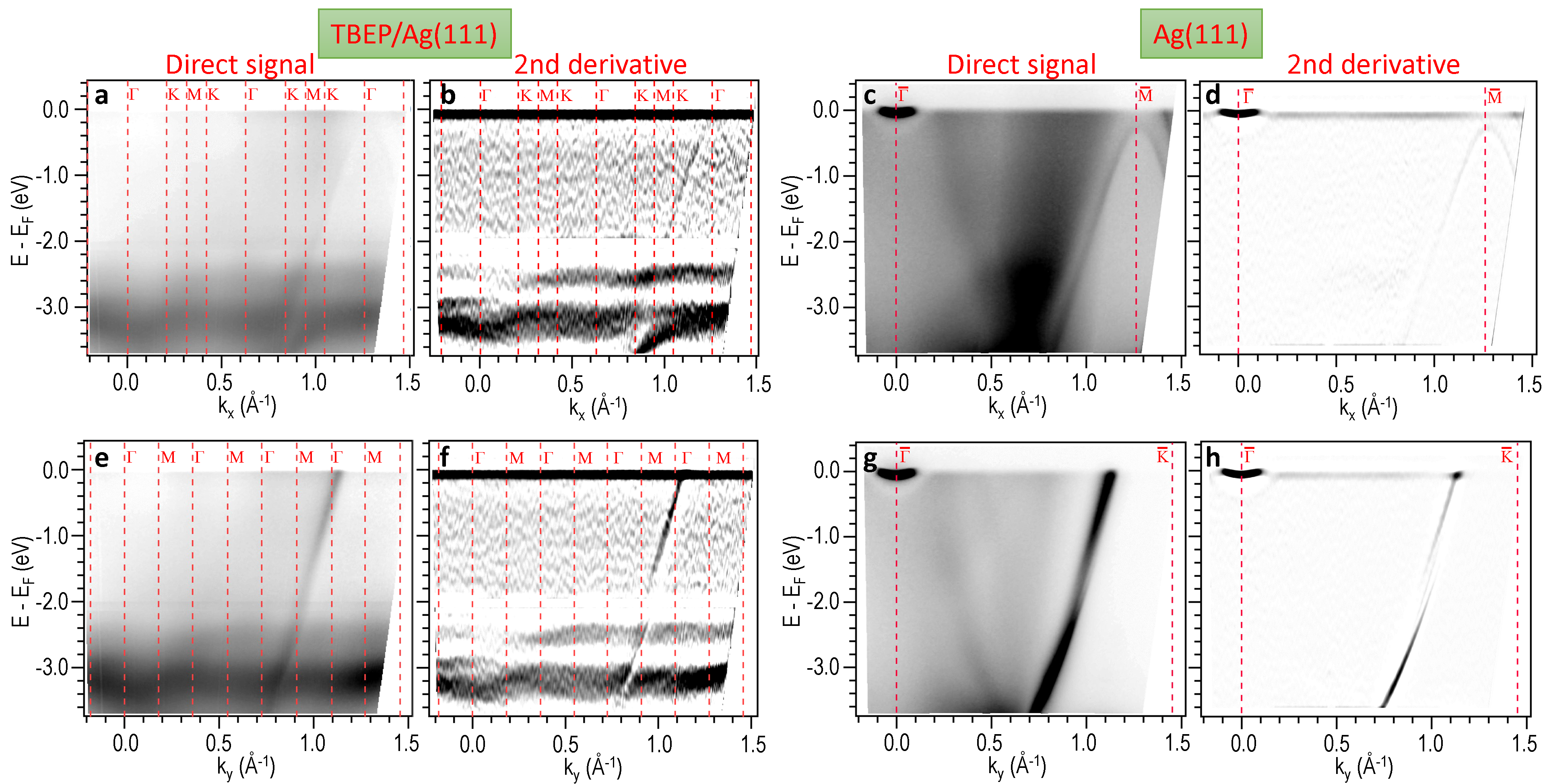}
	\caption*{Figure S3. Electronic band structure comparison of the 2D organometallic network versus the Ag(111) substrate along the two high-symmetry directions. The direct photoemission signal, measured with a photon energy of 21~eV, is shown in panels (a), (c), (e), and (g), while the corresponding second derivative in the side panels (b), (d), (f), and (h). The highly dispersive \textsl{sp} bands of the substrate at large $k_x$ and $k_y$ values are visible in all panels, whereas the relatively flat band-manifolds at the lowest energy region are only visible in the network's signal (left columns). Contrarily, the Shockley state is only visible at $\Gamma$ for the pristine substrate, as it is depleted in the 2D network. The color grayscale is linear (the darker, the more intense) and the vertical dashed lines indicate the high-symmetry points (rotated by 30$^\circ$ with respect to one another).} 
	\label{SI1_ARPES}
\end{figure*}

\begin{figure*}
	\centering
	\includegraphics[width=.8\textwidth,trim=2 2 2 2,clip]{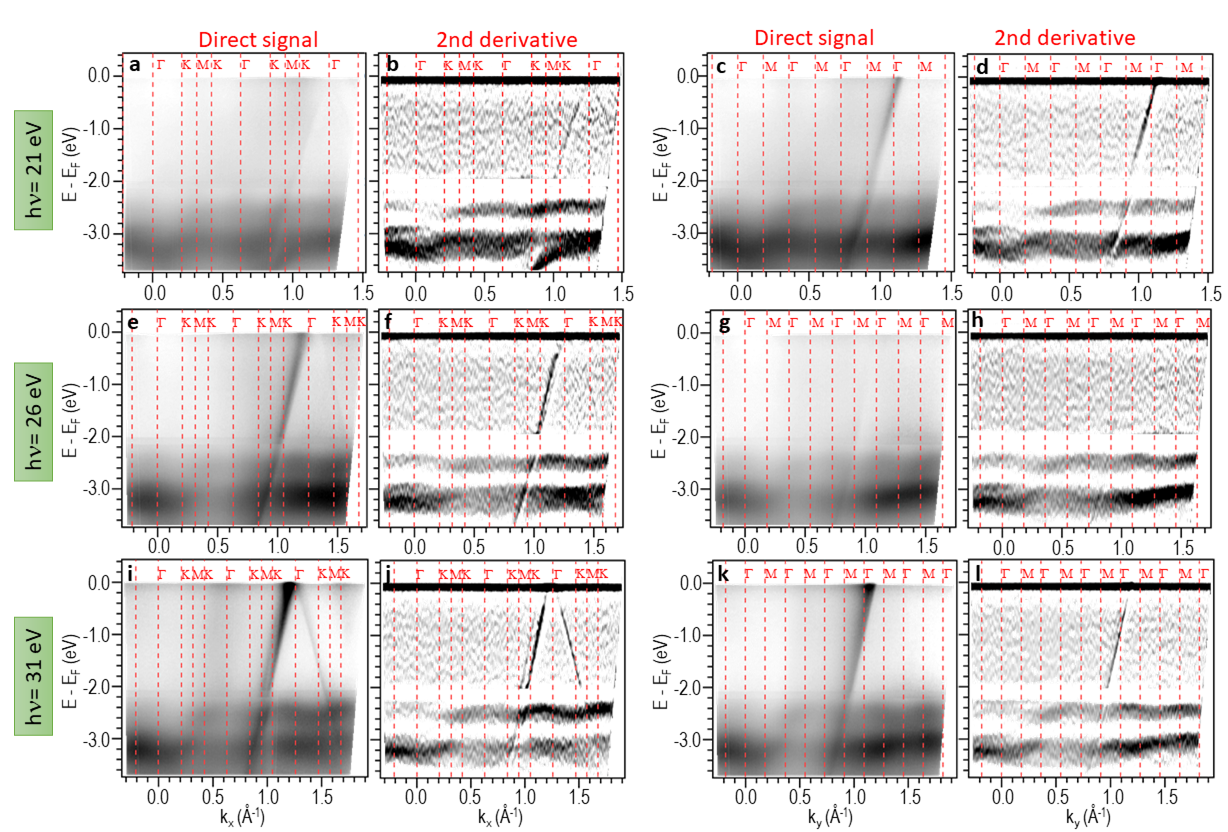}
	\caption*{Figure S4. Electronic band structure evolution of the 2D organometallic network as a function of photon energy along the two high-symmetry directions, specifically at 21~eV (a--d), 26~eV (e--h), and 31~eV (i--l). The direct photoemission signal is shown in panels (a), (c), (e), (g), (i), and (k), while the corresponding second derivative in the side panels (b), (d), (f), (h), (j), and (l). The substrate's \textsl{sp} bands change their dispersion with the photon energy, whereas the band-manifolds of the network do not, corroborating their 2D origin. The color grayscale is linear (the darker, the more intense) and the vertical dashed lines indicate the high-symmetry points (rotated by 30$^\circ$ with respect to one another).}
	\label{SI2_ARPES}
\end{figure*}

\begin{figure*}
	\centering
	\includegraphics[width=\textwidth,trim=2 2 2 2,clip]{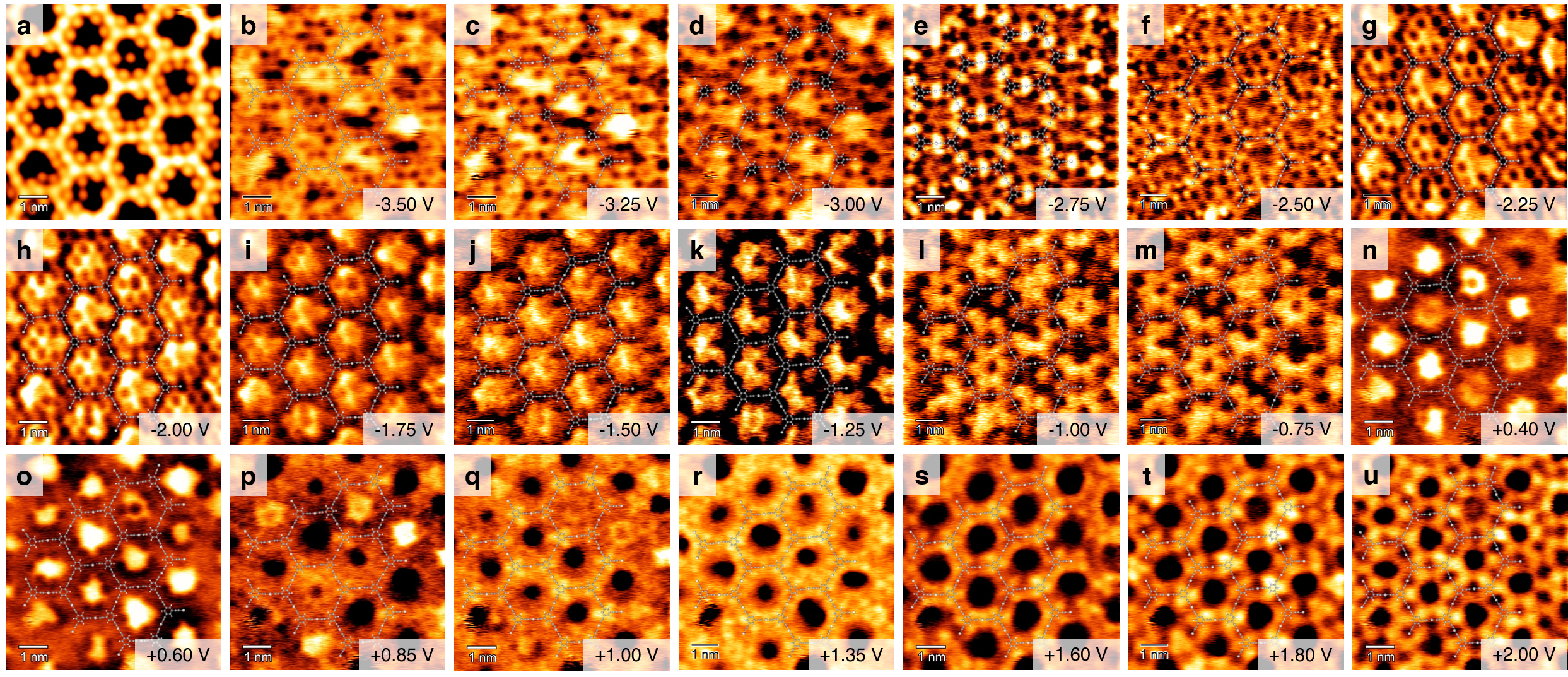}
	\caption*{Figure S5. (a) Reference LT-STM image (STM set-point: $-$2.0~V, 300~pA). (b$-$u) LT \textit{dI/dV} maps taken at different set-point biases with ${V_{\text{RMS}}=10.0}$~mV and ${f_{\text{osc}}=817.3}$~Hz. STS set-points: (b) $-$3.50~V, 450~ pA; (c) $-$3.25~V, 450~pA; (d) $-$3.00~V, 400~pA; (e) $-$2.75~V, 400~pA; (f) $-$2.50~V, 350~pA; (g) $-$2.25~V, 350~pA; (h) $-$2.00~V, 300~pA; (i) $-$1.75~V, 300~pA; (j) $-$1.50~V, 250~pA; (k) $-$1.25~V, 250~pA; (l) $-$1.00~V, 200~pA; (m) $-$0.75~V, 200~pA; (n) $+$0.40~V, 100~pA; (o) $+$0.60~V, 100~pA; (p) $+$0.85~V, 120~pA; (q) $+$1.00~V, 120~pA; (r) $+$1.35~V, 150~pA; (s) $+$1.60~V, 150~pA; (t) $+$1.80~V, 200~pA; (u) $+$2.00~V, 200~pA.}
	\label{didvSI}
\end{figure*}

\begin{figure*}
	\centering
	\includegraphics[width=\textwidth,trim=2 2 2 2,clip]{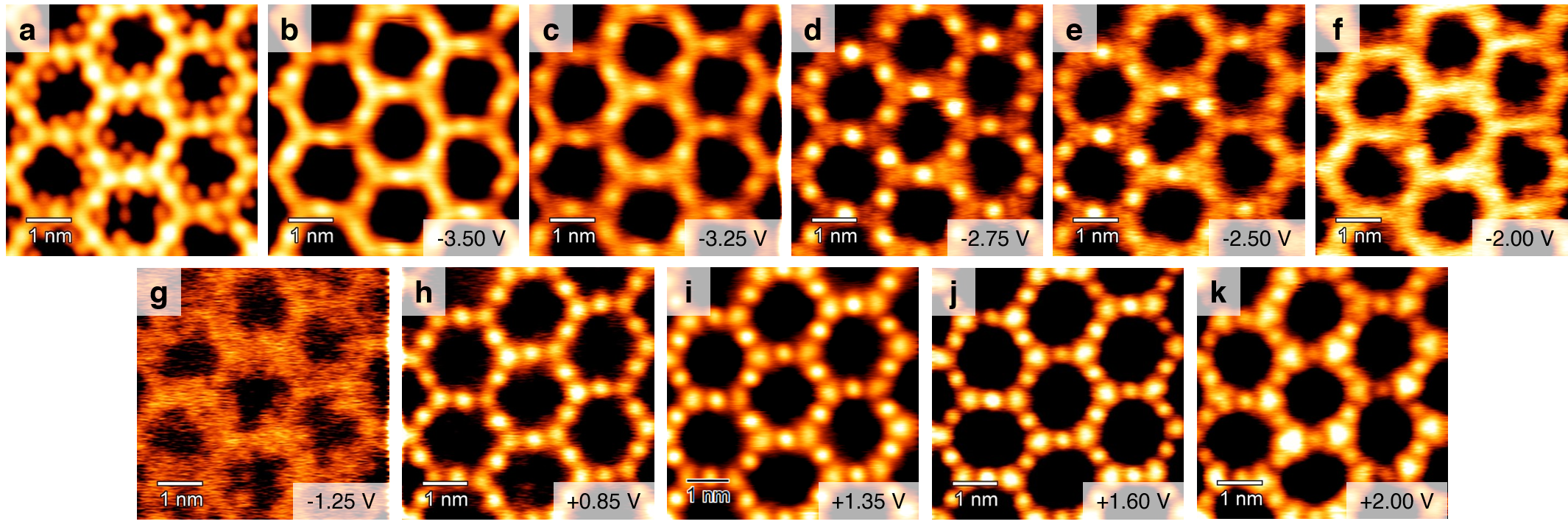}
	\caption*{Figure S6. (a) Reference (constant-current) LT-STM image (STM set-point: $-$100~mV, 50~pA). (b$-$k) Constant-height LT-STM images taken at different set-point biases. STM set-points: (b) $-$3.50~V; (c) $-$3.25~V; (d) $-$2.75~V; (e) $-$2.50~V; (f) $-$2.00~V; (g) $-$1.25~V; (h) $+$0.85~V; (i) $+$1.35~V; (j) $+$1.60~V; (k) $+$2.00~V.}
	\label{CHSI}
\end{figure*}

\begin{figure*}
	\centering
    \includegraphics[width=0.85\textwidth,trim=2 2 2 2,clip]{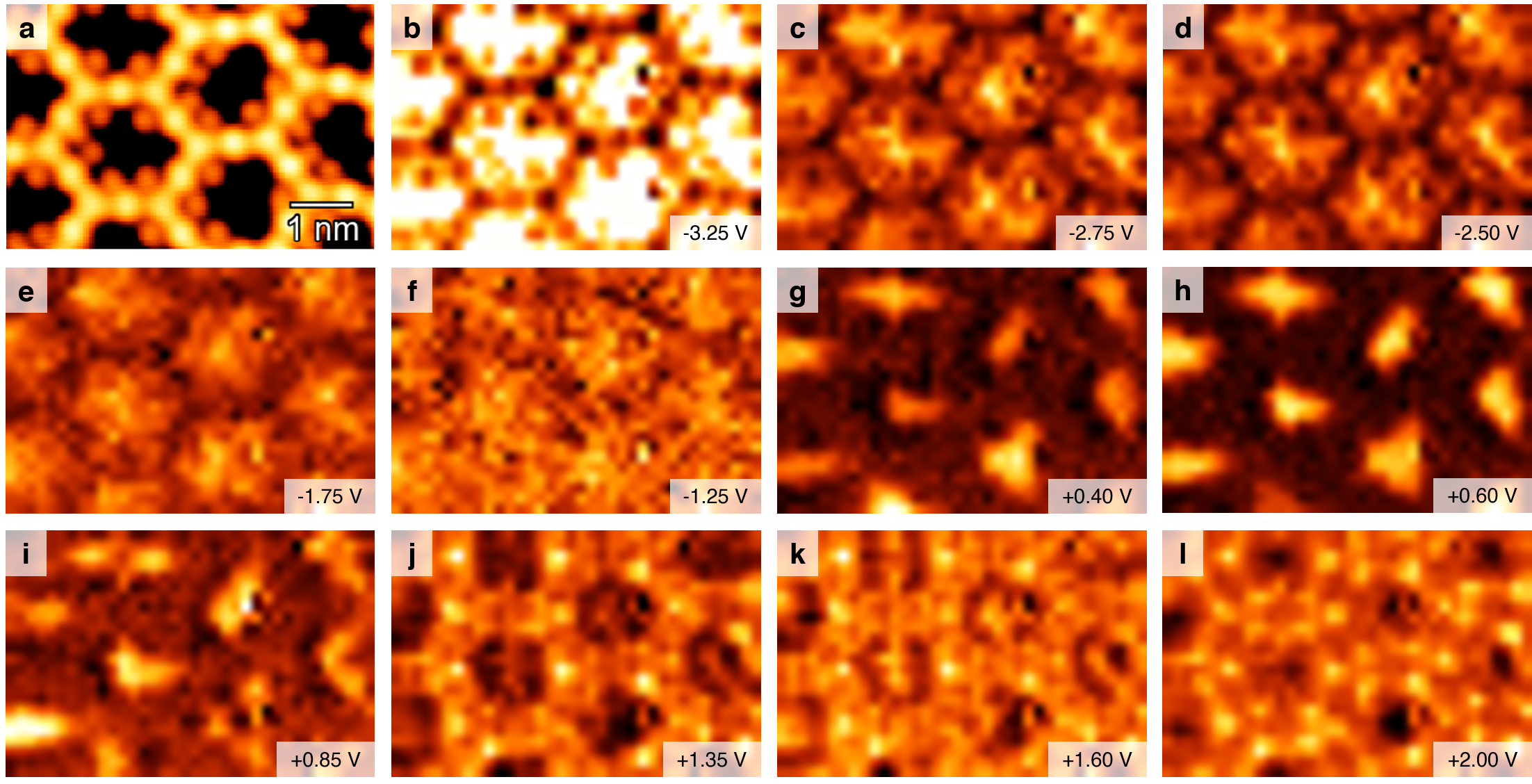}
	\caption*{Figure S7. LT-STS maps of a 36$\times$24 pixel grid collected with ${V_{\text{RMS}}=10.0}$~mV and ${f_{\text{osc}}=817.3}$~Hz. A STS curve is associated with each pixel, and each map is an energy slice at a specific bias voltage. (a) Reference LT-STM image (STM set-point: $-$500~mV, 14~pA). STS set-points: (b) $-$3.50~V; (c) $-$2.75~V; (d) $-$2.50~V; (e) $-$1.75~V; (f) $-$1.25~V; (g) $+$0.40~V; (h) $+$0.60~V; (i) $+$0.85~V; (j) $+$1.35~V; (k) $+$1.60~V; (l) $+$2.00~V.}
	\label{gridSI}
\end{figure*}

\begin{figure*}
	\centering
	\includegraphics[width=.8\textwidth,trim=2 2 2 2,clip]{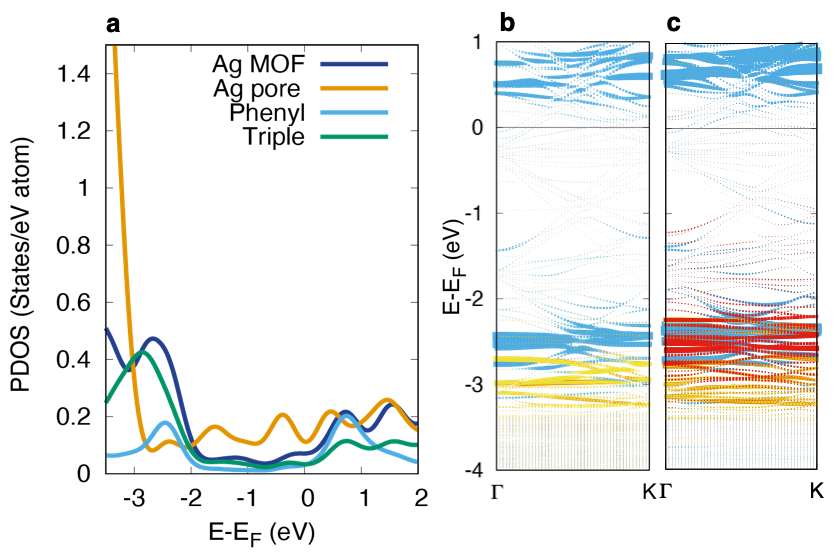}
	\caption*{Figure S8. Theoretical electronic properties of the 2D organometallic network on Ag(111). (a) Density of states at the $\Gamma$ point projected on different groups of atoms. Band dispersion along the high-symmetry path of 2D network without (b) and with (c) Br adatoms in the pores. Carbon bands are reported in blue (\textsl{p$_z$}), and orange/yellow (\textsl{p$_{x,y}$}). Red bands correspond to states mainly localized on Br atoms.}
	\label{theory}
\end{figure*}

\begin{figure*}
	\centering
	\includegraphics[width=.8\textwidth,trim=2 2 2 2,clip]{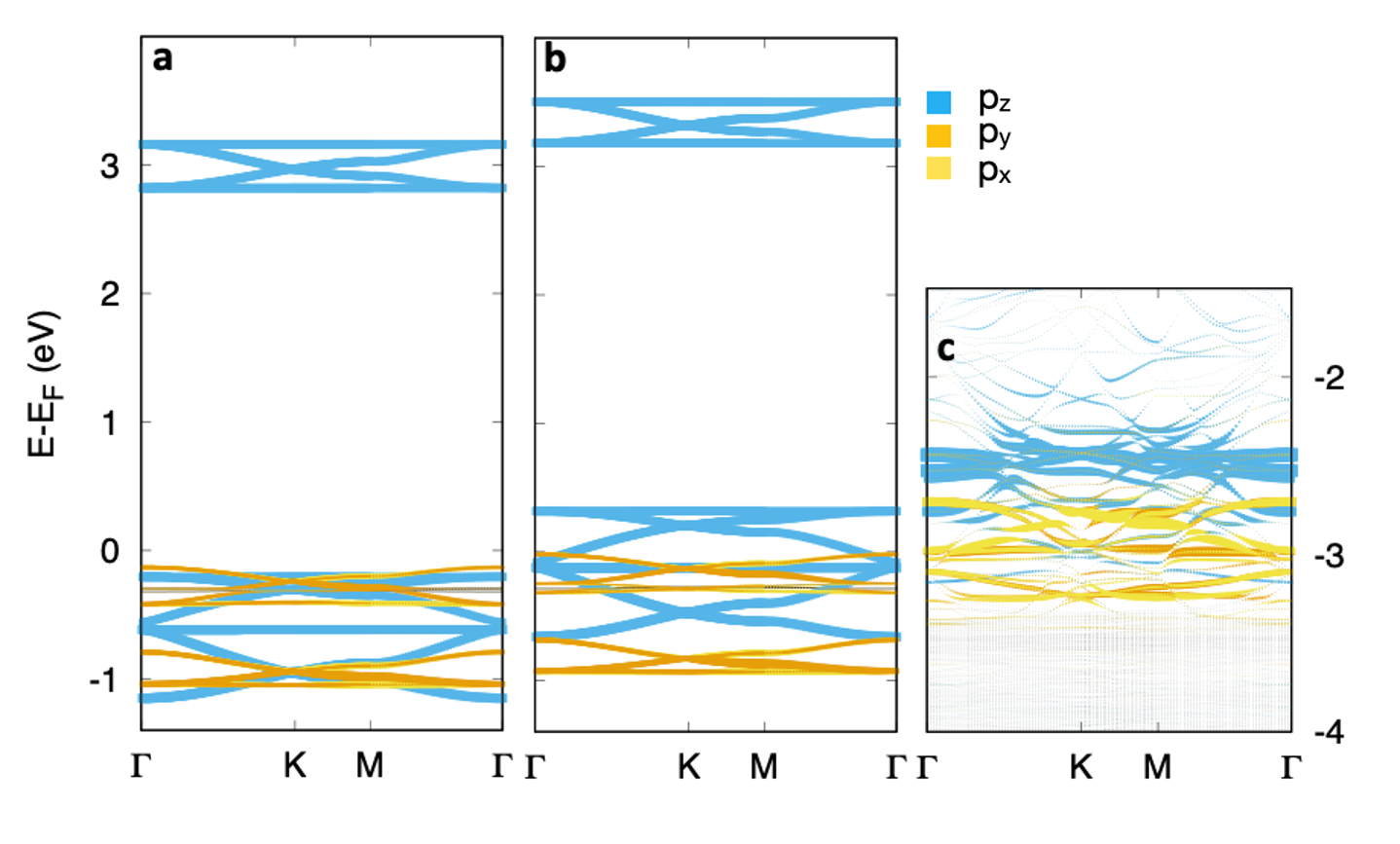}
	\caption*{Figure S9. Majority (a) and minority (b) spin band structure of the free-standing 2D organometallic network along high-symmetry paths of the Brillouin zone. (c) Band structure of the Ag(111)-supported 2D network for comparison. Cyan and yellow/orange lines correspond to different \textsl{p} bands, respectively \textsl{p$_z$} and \textsl{p$_{x,y}$}.}
	\label{calcSTS}
\end{figure*}

\begin{figure*}
	\centering
	\includegraphics[width=.65\textwidth,trim=2 2 2 2,clip]{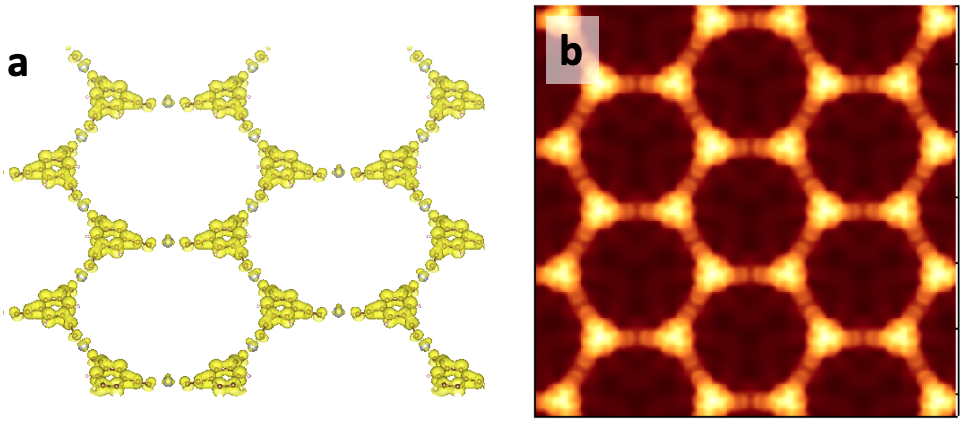}
	\caption*{Figure S10. LDOS (a) and simulated $dI/dV$ map (b) of the Ag(111)-supported 2D network at $+$2.0~eV from the Fermi level.}
	\label{free-standing}
\end{figure*}

\end{document}